\documentclass[%
 reprint, nofootinbib,
 amsmath,amssymb,
 aps
]{revtex4-2}

\usepackage[dvipsnames]{xcolor}

\usepackage{appendix}
\usepackage[utf8]{inputenc}
\usepackage{soul}
\usepackage{graphicx}
\usepackage{mathrsfs}
\usepackage{hyperref}
\usepackage{braket}
\usepackage{amsthm}
\usepackage{float}
\usepackage{notoccite} 
\hypersetup{
    colorlinks=true,
    linkcolor=magenta,
    citecolor = blue,
    filecolor=magenta,      
    urlcolor=blue,
    pdftitle={Overleaf Example}
    }
\usepackage{media9}
\usepackage{float}
\usepackage{dirtytalk}
\usepackage{mathrsfs}
\usepackage{amsmath,amssymb}
\usepackage{physics}
\usepackage{tikz}
\usetikzlibrary{quantikz}
\usepackage[export]{adjustbox}
\usepackage{lipsum} % for dummy text only
\graphicspath{{Figures/}}
\usepackage{subfig}
\pdfoutput = 1
\usepackage{caption}
\usepackage{mdframed}
\newcommand*{\rom}[1]{\expandafter\romannumeral #1}
\begin{document}

\title{GUP Corrections to the Jaynes-Cummings Model}% Force line breaks with \\

\author{Kabir Khanna}
\affiliation{Department of Engineering Design, Indian Institute of Technology Madras, \\Chennai 600 036, India}
\email{kabirkhanna@alumni.iitm.ac.in}

\author{Saurya Das}
\affiliation{Theoretical Physics Group and Quantum Alberta, 
Department of Physics \& Astronomy, University of Lethbridge, 4401 University Drive, Lethbridge Alberta, T1K 3M4, Canada}
\email{saurya.das@uleth.ca}

\begin{abstract}
The Generalized Uncertainty Principle (GUP) is a modification of the Heisenberg's Uncertainty Principle predicted by several theories of quantum gravity. 
In this work, we compute GUP corrections to the well known Jaynes-Cummings Model (JCM) with the aim of eventually observing quantum gravity effects in quantum optical systems. 
To this end, we first analytically solve the GUP-corrected JCM and obtain the corrected Rabi frequency in the quadratic GUP model.
Following this, we calculate the effects of a dispersive interaction with light in a coherent state, and show that this gives rise to  photon-added coherent states that were first studied in \cite{PhysRevA.43.492}.
The latter causes a change in the value of the Wigner function, which if detected in the laboratory, would in effect be a signature of quantum gravity. 

\end{abstract}
\maketitle

\section{Introduction}
A successful theory of quantum gravity (QG), consistent with both general relativity and quantum mechanics and capable of making testable predictions continues to elude us. 
Furthermore, one normally expects quantum gravitational effects to play a role at or near the Planck scale $(\sim10^{16} \text{TeV})$, which is way beyond the maximum energies accessible in accelerators, namely the electroweak scale $(\sim 1 \,\text{TeV})$. 
While this makes it practically impossible to test candidate theories of QG, 
there is no \textit{a priori} reason to assume that no 
QG effects would be present, albeit indirectly, in the $15$ orders of magnitude intervening between the Planck and electroweak scales. \\
A common feature of various QG theories is the existence of a minimal measurable length.
This results in an effective modification of the Heisenberg Uncertainty Principle (HUP) to the so-called the Generalized Uncertainty Principle (GUP), which in turn gives rise to 
QG terms in practically all quantum Hamiltonians. This has given rise to many 
GUP-based phenomenological models, which have been used predict QG effects in low-energy quantum systems \cite{PhysRevD.84.044013, PhysRevLett.101.221301, ALI2009497,osti_22701519, shababi2020non, aghababaei2022quantum, lemos2022quantum,bosso2022bell}. Various experimental tests based on these models have been proposed using quantum opto-mechanical interactions \cite{Pikovski2012Probing, PhysRevA.96.023849}, gravitational bar detectors \cite{Marin:2013pga}, nano-diamond interferometry \cite{PhysRevA.90.033834}, and direct measurements on a macroscopic harmonic oscillator \cite{2015}.\\
In this work, we study the implications of these models on the light-atom interaction using the Jaynes-Cummings Model (JCM) \cite{1443594}. Since its advent in 1963 by Edwin Jaynes and Fred Cummings, the JCM has been adapted and applied to multiple areas of physics. This exactly solvable model forms an important component in quantum optics \cite{gerry_knight_2004}, many-body physics \cite{PhysRevLett.103.086403}, quantum computing \cite{nielsen_chuang_2010, *azuma2011quantum, *PhysRevA.69.062320}, and quantum simulation \cite{2020} (see \cite{greentree2013fifty} for an extensive review of the applications of the JCM). Experimentally, it can be realized most naturally through a cavity quantum electrodynamic or an ion-trap setup. The pervasiveness of this model throughout theory and experiments in quantum mechanics provides the motivation for us to look into its GUP modifications. To the best of our knowledge, this is the first time that this study is being done. 
\\ 
We start with the GUP expressed as a modified commutation relation between position and its canonically conjugate momentum 
of the form 
\cite{PhysRevD.96.066008}
\begin{equation}
    \label{fullgupmodel}
    [\hat{q}, \hat{p}] = \hbar\big[1-2\delta\gamma'\hat{p} + 4\epsilon\gamma'^2\hat{p}^2 \big]~.
\end{equation}
In the above, $(\hat{q}, \hat{p})$ are the usual position and momentum operators, $\gamma' = \gamma_{0}/M_{Pl}c$ with $M_{Pl}$ being the Planck mass, and $\gamma_{0}$ is a free parameter.
Unlike in \cite{bosso2018potential}, 
we do not assume 
$\gamma_{0} \sim 1$, and leave it for experiments to determine its value. $\delta$ and $\epsilon$ are independent parameters that parametrize the magnitude of the linear and quadratic in momentum corrections in eq.(\ref{fullgupmodel}) above. They are typically of $\mathcal{O}(1)$ and produce some well-known GUP models. For instance, $\delta = 0$ and $\epsilon=1/4$ gives rise to the model used in \cite{PhysRevD.52.1108}. Following the quantization of the electromagnetic field, we derive corrections to the JCM Hamiltonian. 
We note here that the effects due to the linear GUP (LGUP) corrections (that typically dominate the quadratic corrections) may be neglected in comparison with the quadratic GUP (QGUP) corrections. As we will discuss further, this happens once we make the Rotating Wave Approximation (RWA). The detailed justification is given in the appendix \ref{appendix:perturbative}. We solve the resultant QGUP JCM and obtain the modified Rabi frequency, which we note is too small to be measured. Following this, we show that a dispersive interaction of the atom with the light field in a coherent state gives rise to photon-added coherent states for time scales comparable to the duration of standard low-energy quantum optical experiments. We then discuss the possibility of probing these states by measuring the change in the Wigner function
demonstrating the
possibility of detecting GUP/QG
effects in the near-future. Towards the end, we summarize our results and discuss implications.\\
This paper is organized as follows: In section \ref{reviewstuff}, we briefly review the quantization of the electromagnetic field and the JCM while setting up the notation from \cite{gerry_knight_2004}. In section \ref{correctionsquadratic}, we compute the Rabi frequency for the QG-corrected JCM. 
In section \ref{largedetuning}, we solve for an interaction with large detuning — a parameter that decides the amplitude and frequency of the Rabi oscillations in the JCM. We then comment on the experimental viability of our results and briefly discuss a way to potentially detect GUP/QG effects. In section \ref{discussion}, we summarize our results and discuss the broader implications of our work.

\section{Review and Setup of the Standard JCM}

\label{reviewstuff}

Following \cite{gerry_knight_2004}, we write the operator form for the electric and magnetic fields of a single mode as 
\begin{align}
\label{elecoperator}
    \hat{E}_{x}(z, t) &= \bigg(\frac{2\omega^2}{V\epsilon_{0}}\bigg)^{1/2}\hat{q}(t)\sin(kz)\\
\label{magneticoperator}
    \hat{B}_{y}(z, t) &= \bigg(\frac{\mu_{0}\epsilon_0}{k}\bigg)\bigg(\frac{2\omega^2}{V\epsilon_{0}}\bigg)^{1/2}\hat{p}(t)\cos(kz)
\end{align}

where $\hat{q}(t)$ and $\hat{p}(t)$ capture the time-dependence of the electric and magnetic fields respectively, $\omega$ is the angular frequency and $k = \omega/c$ is the wave number. The constant $V$ is the ``effective volume" of the cavity in which the field is confined, and $\epsilon_0$ and $\mu_0$ refer to the usual permittivity and permeability constants in vacuum.\\
The Hamiltonian of the system, namely 
\begin{equation}
    H = \frac{1}{2}\int dV \bigg[\epsilon_0E_{x}^2(z,t)+\frac{1}{\mu_0}B_{y}^2(z,t)\bigg],
\end{equation}
can be re-written using Eqs.\eqref{elecoperator} and \eqref{magneticoperator} as

\begin{equation}
    \hat{H} = \frac{1}{2}\big(\hat{p}^2 + \omega^2\hat{q}^2).
\end{equation}
  
In the case of the quantum harmonic oscillator, $\hat{q}$ and $\hat{p}$ are the canonical position and momentum operators where normally (in the absence of QG corrections), the standard commutation relation $[\hat{q}, \hat{p}] = i\hbar$ is used. Here, the electric and magnetic fields play the role of canonical position and momentum respectively. 
Analogous to the harmonic oscillator, one then defines annihilation $(\hat{a})$ and creation $(\hat{a}^{\dag})$ operators as follows\footnote{We omit the subscript $k$ signifying the mode in $a_{k}(a_k^{\dag})$ since we work with only a single-mode of light}
\begin{align}
    \label{annihilationcanonical}
    \hat{a} &= \sqrt{\frac{1}{2\hbar\omega}}(\omega \hat{q} + i\hat{p})\\
    \label{creationcanonical}
    \hat{a}^{\dag} &= \sqrt{\frac{1}{2\hbar\omega}}(\omega \hat{q} - i\hat{p})
\end{align}
such that $[\hat{a}, \hat{a}^{\dag}] = 1$. Consequently, the electric and  magnetic field operators and the Hamiltonian become
\begin{align}
    \label{elecfieldaadag}
    \hat{E}_{x}(z, t) &= \mathcal{E}_{0}(\hat{a}+\hat{a}^{\dag})\sin(kz)\\
    \label{bfieldaadag}
    \hat{B}_{y}(z, t) &= \frac{\mathcal{B}_{0}}{i}(\hat{a}-\hat{a}^{\dag})\cos(kz)\\
    \label{hamiltonianaadag}
    \hat{H} &= \hbar\omega\bigg(\hat{a}^{\dag}\hat{a} + \frac{1}{2}\bigg)
\end{align}
where $\mathcal{E}_{0} =(\hbar\omega/\epsilon_0V)^{1/2}$ and $\mathcal{B}_{0} =(\mu_0/k)(\epsilon_0\hbar\omega^3/V)^{1/2}$ signify the electric and magnetic field per quanta. \\
As is well-known, the JCM is a fully quantum mechanical model of interaction of a single-mode of light with a two-level atom based on the above formalism 
(see \cite{doi:10.1080/09500349314551321} for a review of the model).
We review the basic formalism and notations here. 
Let us consider a two-level atom with levels $\ket{e}$ and $\ket{g}$ interacting with a single-mode light field of the form \eqref{elecfieldaadag}, except now
we consider an arbitrary polarization vector $\mathbf{e}$ instead of one 
along the $x$ direction.
The JCM interaction Hamiltonian is then given by a simple dipole term
\begin{equation}
    \label{interactionhamiltonian}
    \hat{H}_I = -\hat{\mathbf{d}}.\hat{\mathbf{E}} = \hat{d}g(\hat{a}+\hat{a}^{\dag})
\end{equation}
where $\hat{d} = \hat{\mathbf{d}}.\mathbf{e}$ and $g = -(\hbar\omega/\epsilon_{0}V)^{1/2}\sin(kz)$.\\
Following \cite{gerry_knight_2004}, we can introduce the atomic transition operators
\begin{equation}
    \hat{\sigma}_{+} = \ket{e}\bra{g},\ \hat{\sigma}_{-} = \ket{g}\bra{e}
\end{equation}
and the inversion operator, 
\begin{equation}
    \hat{\sigma}_{3} = \ket{e}\bra{e} -\ket{g}\bra{g}.
\end{equation}
It is easy to check that $[ \hat{\sigma}_{+},  \hat{\sigma}_{-}] =  \hat{\sigma}_{3}$ and $[ \hat{\sigma}_{3},  \hat{\sigma}_{\pm}] =  2\hat{\sigma}_{\pm}$. Furthermore, from parity considerations, we see that the diagonal elements of the dipole operator $\bra{e}\hat{d}\ket{e} = \bra{g}\hat{d}\ket{g} = 0$, implying
\begin{align}
    \hat{d} &= d\ket{g}\bra{e} + d^{*}\ket{e}\bra{g}\\
    &=d(\hat{\sigma}_{-} + \hat{\sigma}_{+})
\end{align}
where we write $\bra{e}\hat{d}\ket{g} = d$, which can be assumed to be real without loss of generality \cite{gerry_knight_2004}.
Using the above results, 
we can finally write the total Hamiltonian for the JCM as
\begin{align}
    \label{jcmparts}
    \hat{H} &= \hat{H}_{A} + \hat{H}_{F} + \hat{H}_{I} \\
     &= \frac{1}{2}\hbar\omega_0\hat{\sigma}_{3} + \hbar{\omega}\hat{a}^{\dag}\hat{a} + \hbar\lambda(\hat{\sigma}_{+} + \hat{\sigma}_{-})(\hat{a}+ \hat{a}^{\dag})
\end{align}
where $\hat{H}_A$ is the free atomic Hamiltonian with energy levels $E_{e}$ and $E_{g}$, $\hat{H}_F$ is the free-field Hamiltonian, and $\lambda = dg/\hbar$ captures the interaction strength between the atom and the field. The energy levels of the atom are $E_{e}=-E_{g} = (1/2)\hbar\omega_0$.\\

\section{Corrections to the JCM and Solution with QGUP}
\label{correctionsquadratic}
In this section, we consider the GUP model 
defined by equation \eqref{fullgupmodel} and compute corrections to the Hamiltonian and the Rabi frequency for the interaction around resonance. Note that the dimension of $\hat{p}$ (from \eqref{annihilationcanonical}) is $\sqrt{\text{Energy}}$ whereas that of $\hat{q}$ (from \eqref{creationcanonical}) is $\sqrt{\text{Energy}}\times\text{Time}$. In order to impose GUP on the quantized electric and magnetic fields, which play the role of $\hat{q}$ and $\hat{p}$ respectively, we take the following ansatz from 
\cite{bosso2018potential}
(with redefined constants),  
\begin{equation}
    \label{finalgup}
    [q, p] = i\hbar\,(1-2\delta\gamma p + 4\epsilon\gamma^{2}p^{2})
\end{equation}
where\footnote{$\gamma$ here was denoted as $\gamma_{EM}$ in \cite{bosso2018potential}.}
\begin{equation}
    \gamma = \frac{\gamma_{0}}{\sqrt{M_{Pl}}c}
    \label{gamma1}
\end{equation}
and $\gamma_{0}$ is a free dimensionless parameter whose values may be fixed through experiments. It can be seen from Eqs.(\ref{finalgup}) and (\ref{gamma1}) that the GUP terms become important when ${\gamma_{0}}p/{\sqrt{M_{Pl}}}c\, \simeq 1$. Associating $p$ with an energy scale $E$ and length scale $L$ such that $p\simeq \sqrt{E}$ (from equations \eqref{annihilationcanonical} and \eqref{creationcanonical}) and $E\simeq \hbar c/L$, we see that 
\begin{equation}
\label{lengthscale}
    \frac{\gamma_0p}{\sqrt{M_{Pl}c}} \simeq 1 \implies \sqrt{\frac{\gamma_0^2l_{Pl}}{L}}\simeq 1,
\end{equation}
where we have used $\sqrt{M_{Pl}}c = \sqrt{l_{Pl}/\hbar c}$. The above equation thereby sets a length scale for the GUP in \eqref{finalgup} as $\gamma_0^2l_{Pl}$. The energy scale $E$ can be associated with the frequency $\omega$ of the electromagnetic wave. Consequently, the length scale $L$ is set by the wavelength of the electromagnetic wave where $\omega$ and $L$ can be related with $\omega\simeq2\pi c/L$.\\
Since this length scale $L$ has not been observed in any experiment as yet, 
the best upper bounds can be imposed on it from the electroweak length scale 
$l_{w}\sim 10^{-18}\text{ m}$, which is the minimum length scale probed so far by any (high energy) experiment. \\
This length scale naturally sets the bound $\gamma_{0}$ $\leq 10^{8}$ (using \eqref{lengthscale}). Additionally, note that $\gamma = \gamma_{0}/(M^{1/2}_{p}c)= \gamma_{0}/(4.4\times10^4)$ in SI units. This sets the bound on $\gamma$ as $\gamma\leq 10^3$. On the other hand, if we take $\gamma_{0}=1$ assuming the perturbation to be effective at the Planck-scale,  this would imply a lower bound on $\gamma$ as well with 
$
\mathcal{O}(
\gamma
)
\geq10^{-5}$. These bounds will help us estimate bounds on the measurable parameters in the coming sections. \\
With the GUP defined in \eqref{finalgup}, we can use the results from \cite{PhysRevD.96.066008} to study the corrections to the JCM. From \cite{PhysRevD.96.066008}, the GUP modified free light-field Hamiltonian has eigenstates that we denote by $\ket{\Tilde{n}}$. The GUP altered raising and lowering operators are denoted by $\Tilde{a} \text{ and } \Tilde{a}^{\dag}$. Their action on the eigenstates $\ket{\tilde{n}}$ is same as that of the standard $\hat{a}$ and $\hat{a}^{\dag}$ on $\ket{n}$.
Additionally, the operator $\hat{q}$ has a GUP dependent correction when written in terms of $\tilde{a}$ and $\tilde{a}^{\dag}$. Using this we write the modified electric field as
\begin{multline}
    \hat{\textbf{E}} = \textbf{e}\bigg(\frac{2\omega^2}{V\epsilon_{0}}\bigg)^{1/2}\sin(kz)\bigg[(\tilde{a} + \tilde{a}^{\dag})\sqrt{\frac{\hbar}{2\omega}}-2i(\tilde{a}^{\dag 2} - \tilde{a}^{2})\frac{\delta\hbar\gamma}{2}\\ -\{(\tilde{a}^{3} + \tilde{a}^{\dag 3})(2\delta^2 + \epsilon) + (\tilde{a}\tilde{N} + \tilde{N}\tilde{a}^{\dag})(3\delta^{2}-2\epsilon)\}\hbar\sqrt{\frac{\hbar\omega}{2}}\gamma^2\bigg].
\end{multline}
Following similar steps as from \eqref{interactionhamiltonian} to obtain the interaction term in \eqref{jcmparts}, the above electric field alters the interaction term introduced in \eqref{interactionhamiltonian} to 
\begin{multline}
    \label{modifiedinteractionhamiltonian}
    \hat{H}_{I} = \hbar\lambda(\hat{\sigma}_{-} + \hat{\sigma}_{+}) \bigg[(\tilde{a} + \tilde{a}^{\dag})-i\sqrt{2\hbar\omega}\delta\gamma(\tilde{a}^{\dag 2} - \tilde{a}^{2})\\ -\hbar\omega\gamma^2\{(\tilde{a}^{3} + \tilde{a}^{\dag 3})(2\delta^2 + \epsilon) + (\tilde{a}\tilde{N} + \tilde{N}\tilde{a}^{\dag})(3\delta^{2}-2\epsilon)\}\bigg]
\end{multline}
where $\lambda = dg/\hbar$, and $\hat{\sigma}_{\pm}$ are the atomic transition operators defined in the previous section.

Next, we make the Rotating Wave Approximation (RWA) \cite{gerry_knight_2004}. This means that certain terms which vary rapidly can be neglected as they average out to 0 around resonance $(\omega \approx \omega_{0})$. Specifically, 
we consider the terms
\begin{align}
    \label{w0-w}
    \hat{\sigma}_{+}\tilde{a} \sim e^{it(\omega_{0}-\omega)}\\
    \label{-w0-w}
     \hat{\sigma}_{-}\tilde{a}^{\dag} \sim e^{-it(\omega_{0}-\omega)}\\
     \label{w0+w}
    \hat{\sigma}_{+}\tilde{a}^{\dag} \sim e^{it(\omega_{0}+\omega)}\\
    \label{-w0+w}
    \hat{\sigma}_{-}\tilde{a} \sim e^{-it(\omega_{0}+\omega)} ~.  
\end{align}
Since the last two terms vary rapidly in comparison to the first two, they can be
safely neglected. Another way to see this is by integrating the time-dependent Schrödinger equation which leads to terms with factors $(\omega_{0}+\omega)^{-1}$ and $(\omega_{0}-\omega)^{-1}$. Around resonance, the term with the factor of $(\omega_{0}+\omega)^{-1}$ is neglected.\\
In this approximation, we can additionally neglect terms from the Hamiltonian \eqref{modifiedinteractionhamiltonian} that are quadratic and cubic in $\tilde{a}\;(\tilde{a}^{\dag})$ in addition to the already neglected terms in equation \eqref{w0+w} and \eqref{-w0+w}. With this, we see that only the last term of the Hamiltonian \eqref{modifiedinteractionhamiltonian} with the QGUP parameter $\gamma^2$ remains and LGUP corrections can be neglected. 

However, note that the parameter $\gamma$ can be small, in which case two things need to be clarified. First, the contribution from terms \eqref{w0+w} and \eqref{-w0+w} is not small compared to terms in the Hamiltonian that contain the GUP factors $\gamma$, and thereby cannot be neglected under the RWA. Secondly, it is not clear whether the smallness 
that results from the rapidly varying factor $\hat{\sigma}_{+}\tilde{a}^2\;(\hat{\sigma}_{-}\tilde{a}^{\dag2})$ is sufficient to negate the smallness due to the extra $\gamma$ factor in the last term with the slowly varying factor $\hat{\sigma}_{+}\tilde{a}\tilde{N}\; (\hat{\sigma}_{-}\tilde{N}\tilde{a}^{\dag})$ from \eqref{modifiedinteractionhamiltonian}. 
Noting these potential two caveats, we
continue with this application of the RWA and neglect the terms that are quadratic and cubic in $\tilde{a}\;(\tilde{a}^{\dag})$, in addition to the terms \eqref{w0+w} and \eqref{-w0+w}, and fully justify our approximation in appendix \ref{appendix:perturbative}, where we show that there exist reasonable experimental regimes for which this approximation can clearly be made. 
Finally, note that the other terms with the $\gamma^2$ factor such as $\sigma_{+}\tilde{a}^3$ vary rapidly, give a tiny contribution in comparison to the other terms and can be safely neglected.\\
Next, in addition to the above, we see that the free light-field Hamiltonian is also modified under GUP \cite{PhysRevD.96.066008} as 
\begin{equation}
        \hat{H}_{F} = \hbar\omega\bigg[\bigg(\tilde{N}+\frac{1}{2}\bigg)-\frac{\hbar\omega}{2}\gamma^{2}\{4(\tilde{N}^{2}+\tilde{N})(\delta^{2}-\epsilon) + \delta^2-2\epsilon\}\bigg].
\end{equation}
This completes the GUP corrections we wish to incorporate in the Hamiltonian. After the modification of electromagnetic field dependent parts \eqref{jcmparts} of the Hamiltonian and making the RWA, the JCM Hamiltonian can be written as
\begin{multline}
\label{fullhamiltonian}
    \hat{H} =  \frac{1}{2}\hbar\omega_{0}\hat{\sigma}_{3} + \hbar\omega\big[\tilde{N}-\{4(\tilde{N}^{2}+\tilde{N})\chi + \beta\}  \big] + \\\hbar\lambda\{\hat{\sigma}_{+}(\tilde{a}-\tilde{a}\tilde{N}\phi) + \hat{\sigma}_{-}(\tilde{a}^{\dag} - \tilde{a}^{\dag}(\tilde{N}+1)\phi)\}
\end{multline}
where 
\begin{gather}
    \phi=\hbar\omega\gamma^{2}(3\delta^{2}-2\epsilon)\\
    \chi=(\hbar\omega\gamma^{2}/2)(\delta^{2}-\epsilon)\\
    \beta =(\hbar\omega\gamma^{2}/2)(\delta^{2}-2\epsilon)
\end{gather}
with $8\chi = \phi + 2\beta$ \footnote{Strictly speaking, these parameters are not independent. We introduce them to make it easier to work with, and give the Hamiltonian a more useful form.}.

With the above Hamiltonian in place, we now proceed to solve 
it. Since there are only linear terms in $\tilde{a}(\tilde{a}^{\dag})$ modulo factors of $\tilde{N}$, the subspace of possible state vectors is restricted to $\ket{e, \tilde{n}}$ and $\ket{g, \tilde{n+1}}$. Thus, we begin by considering a general state of the atom-light field as

\begin{equation}
\label{ansatzquadratic}
    \ket{\psi(t)} = C_{e,n}(t)\ket{e, \tilde{n}} + C_{g,n+1}(t)\ket{g, \tilde{n+1}} 
\end{equation}
with the initial state as $\ket{\psi(0)} = \ket{e, \tilde{n}}$. By solving the Schrödinger equation for the above system assuming $\Delta =\omega_0 - \omega \approx 0$, we evaluate $C_{e,n}(t)$ and $C_{g,n+1}(t)$ up to $\mathcal{O}(\phi)(\mathcal{O}(\gamma^2))$ as

\begin{align}
\label{solutionen}
 &C_{e.n}(t) = \cos(\Omega_{QG}(n)\;\text{t})\bigg[1-2(n+1)\phi-4\sqrt{n+1}\;\chi\frac{\omega}{\lambda}\bigg] \\
 \label{solutiongn}
 &C_{g,n+1}(t) = -i\sin(\Omega_{QG}(n)\;\text{t})\big[1-2(n+1)\phi\big]
\end{align}
where 
\begin{gather}
\begin{split}
    \Omega_{QG}(n) = \lambda\sqrt{n+1}\big[1-(n+1)\phi\big]. 
    \end{split}
\end{gather}
Note that for $\chi=\phi=\gamma=0$, \eqref{solutionen} and \eqref{solutiongn} give the solutions of the standard JCM. With the GUP modified JCM solved, we can now calculate that atomic inversion as
\begin{align}
   W(t) &= \abs{C_{e,n}}^2 -  \abs{C_{g,n+1}}^2\\
   & = \cos(2\Omega_{QG}(n)\text{t}) \;\;\;\;\; (\text{upto } \mathcal{O}(\gamma^2))
\end{align}
which gives the Rabi frequency as 
\begin{equation}
    \label{correctedrabiq}
    \Omega_{QG}(n) = \Omega(n)(1-(n+1)\phi) 
\end{equation}

where $\Omega(n) = 2\lambda\sqrt{n+1}$ is the standard Rabi frequency without the GUP corrections, obtained when we set the GUP dependent term $\phi=0$ in the above. For a GUP perturbation at the electroweak scale, $\omega = 10^{16}$ Hz, and ${n} = 1$, the change in the Rabi frequency due to GUP effects is $\sim 10^{-12}$ Hz. This change is extremely small to be measured. Nevertheless, we now proceed to study a different variation of the JCM, in which we argue that experimental detection of GUP effects is potentially achievable in the near future.
% \newpage
% \vfill\null
\section{QGUP Corrected JCM with Large Detuning}
\label{largedetuning}
We consider the case of large detuning in the QGUP corrected JCM. In the standard JCM, this is of interest since it allows for the creation of macroscopically distinguishable states which are important in many fundamental tests of quantum mechanics \cite{brune1994lamb, guo1996generation}. When GUP is taken into account, the hope is that this exercise would not only provide more insight into the quantum mechanical effects of gravity, but also provide the experimental means to test it.\\
We first derive an effective Hamiltonian $\hat{H}_{\text{eff}}$ for the case of large detuning using Hamiltonian \eqref{fullhamiltonian} as the starting point. 
We arrive at the following effective Hamiltonian as (see appendix \ref{appendix:effectivaction} for details) 
\begin{equation}
\label{effectiveh}
    \hat{H}_{\text{eff}} = \frac{\hbar\lambda^2}{\Delta}[\hat{A},\hat{A}^{\dag}]
\end{equation}
where $\hat{A} = \hat{\sigma}_{+}\tilde{a}(\hat{I} - \tilde{N}\phi)$, $\hat{A}^{\dag} = \hat{\sigma}_{-}\tilde{a}^{\dag}(\hat{I} - (\tilde{N}+\hat{I})\phi)$, and the other operators are as defined in the previous sections. Evaluating $[\hat{A},\hat{A}^{\dag}]$
in the above equation gives the effective Hamiltonian to be
\begin{equation}
    \label{effectivehamiltonian}
    \hat{H}_{\text{eff}} = \hbar\mu\bigg( \hat{\sigma}_{3}(\tilde{N}-2\tilde{N}^2\phi) + \hat{\sigma}_{+}\hat{\sigma}_{-}(\tilde{I} - 2\phi - 4\tilde{N}\phi)\bigg)
\end{equation}
with $\mu = \lambda^2/\Delta$ having dimensions of frequency. Note that although the GUP is not directly needed for going from equation \eqref{effectiveh} to \eqref{effectivehamiltonian}, we do need it for the derivation of \eqref{effectiveh} itself. It is now easy to compute how the effective Hamiltonian evolves the states $\ket{e}\ket{\tilde{n}}$ and $\ket{g}\ket{\tilde{n+1}}$. 
As in the standard JCM case, 
these do not produce any interesting effects — just overall phase factors. 
We begin by considering the simplest possible initial state of the light-atom system with the light field in a coherent state, namely, $\ket{\psi(0)}=\ket{g}\ket{\tilde{\alpha}}$ where $\ket{\tilde{\alpha}}$ are GUP modified coherent states as defined in \cite{PhysRevD.96.066008}. At a time $t>0$, this state evolves as
\begin{align}
    \ket{\psi(t)} &= e^{-i\hat{H}_{\text{eff}}t/\hbar}\ket{\psi(0)}\\
    \label{beforetayloragarwal}
    & = e^{-\abs{\alpha}^2/2} \sum_{n=0}^{\infty}\frac{\alpha^n}{\sqrt{n!}}e^{i\mu t(n-2n^2\phi)}\ket{g}\ket{\tilde{n}}\\
    & = e^{-\abs{\alpha}^2/2} \sum_{n=0}^{\infty}\frac{(\alpha e^{i\mu t})^n}{\sqrt{n!}}e^{-2i\mu t n^2\phi}\ket{g}\ket{\tilde{n}}.
\end{align}
In case of  $\langle(\tilde{a}^{\dag}\tilde{a})^2\rangle2\phi\mu t \ll 1$, $e^{-2i\mu t n^2\phi}$ can be suitably Taylor expanded to $\mathcal{O}(\phi)$. 
Assuming $\mathcal{O}(\abs{\alpha}) = 1$, this can be done as long as $t\ll1/(\phi \mu)$. For $\mu=10^5$ and $\omega = 10^{14}$ Hz, we require $t\ll10^{24}s$ when the GUP perturbation is at Planck scale, and $t\ll10^{8}s$ if the GUP perturbations start taking effect at the electroweak scale. In either case, these times are much higher than the duration for which the experiments are carried out. Thus, we can safely work with the linear term of the Taylor expansion.  
Expanding the above equation up to $\mathcal{O}(\phi)$ using $n\ket{\tilde{n}} = \tilde{a}^{\dag}\tilde{a}\ket{\tilde{n}}$ and $\tilde{a}\ket{\tilde{\alpha}} = \alpha\ket{\tilde{\alpha}}$, we obtain
\begin{multline}
    \label{beforeagarwal}
     \ket{\psi(t)}   = \frac{1}{\mathcal{N}}\bigg(\ket{g}\ket{\tilde{\alpha} e^{i\mu t}} - i2\alpha\phi\mu t e^{i\mu t}(\tilde{a}^\dag\\+\alpha e^{i\mu t} \tilde{a}^{\dag 2})\ket{g}\ket{\tilde{\alpha} e^{i\mu t}}
     \bigg),
\end{multline}
where ${\mathcal{N}}$ is a normalization factor we have put in by hand. 
The above state has a part where the creation operator operates on a coherent state. Accordingly, the resultant states are analogous to the photon added coherent states that were first proposed by Agarwal and Tara in 1991 \cite{PhysRevA.43.492}. For $m$ photon additions to a coherent state, the resultant photon added coherent state is denoted as $\ket{\alpha,m}$ with the appropriate normalization factor. Since we have GUP corrected operators and states in our case, we can similarly define GUP modified photon-added coherent states $\ket{\tilde{\alpha}, \tilde{m}} $ as
\begin{equation}
\label{agarwalgn}
    \ket{\tilde{\alpha}, \tilde{m}} = \frac{\tilde{a}^{\dag m}\ket{\tilde{\alpha}}}{(\bra{\tilde{\alpha}}\tilde{a}^m\tilde{a}^{\dag m}\ket{\tilde{\alpha}})^{1/2}}
\end{equation}
where $\bra{\tilde{\alpha}}\tilde{a}^m\tilde{a}^{\dag m}\ket{\tilde{\alpha}} = L_{m}(-\abs{\alpha}^2)m!$, $L_{m}(x)$ being the Laguerre polynomial of order $m$. With the above definition, equation \eqref{beforeagarwal} can be written as

\begin{multline}
    \label{gcoherent}
    \ket{\psi(t)} = \frac{1}{\mathcal{N}}\bigg(\ket{g}\ket{\tilde{\alpha} e^{i\mu t}} - i2\alpha\phi\mu t\\ \cross\big(e^{i\mu t}k_{\alpha,1}\ket{g}\ket{\tilde{\alpha}e^{i\mu t}, \tilde{1}}+e^{2i\mu t}\alpha k_{\alpha,2}\ket{g}\ket{\tilde{\alpha} e^{i\mu t}, \tilde{2}}\big)\bigg),
\end{multline}
where we have used the notation $k_{\alpha,m} = (L_{m}(-\abs{\alpha}^2)m!)^{1/2} $. Similarly, for the initial state of $\ket{\psi(0)} = \ket{e}\ket{\tilde{\alpha}}$, $\ket{\psi(t)}$ for $t>0$ becomes
\begin{multline}
    \label{ecoherent}
    \ket{\psi(t)} = \frac{1}{\mathcal{N}}\bigg(e^{-i\mu t}(1+i2\phi\mu t)\ket{e}\ket{\tilde{\alpha}e^{-i\mu t}}\\ +i2\phi\mu\alpha t\big( 3k_{\alpha,1}e^{-2i\mu t}\ket{e}\ket{\tilde{\alpha}e^{-i\mu t}, \tilde{1}}\\ + \alpha k_{\alpha,2}e^{-3i\mu t}\ket{e}\ket{\tilde{\alpha}e^{-i\mu t}, \tilde{2}}\big)\bigg).
\end{multline}

The above expressions \eqref{gcoherent} and \eqref{ecoherent} constitute the key results of our work. They suggest that the effect of the GUP on coherent states is that of a multi-photon medium giving rise to photon added or excited coherent states. This photon addition to the coherent states in \eqref{gcoherent} and \eqref{ecoherent} changes the quasi-classical coherent state into a state with non-classical properties. The smaller the amplitude of the initial coherent state $\abs{\alpha}$, the higher the non-classicality of the excited coherent state $\ket{\tilde{\alpha},\tilde{1}}$. Furthermore, these states do not exhibit gaussian statistics in their field quadratures and are termed as non-gaussian states. 

\begin{figure}
    \centering
    \includegraphics[width = 0.5\textwidth]{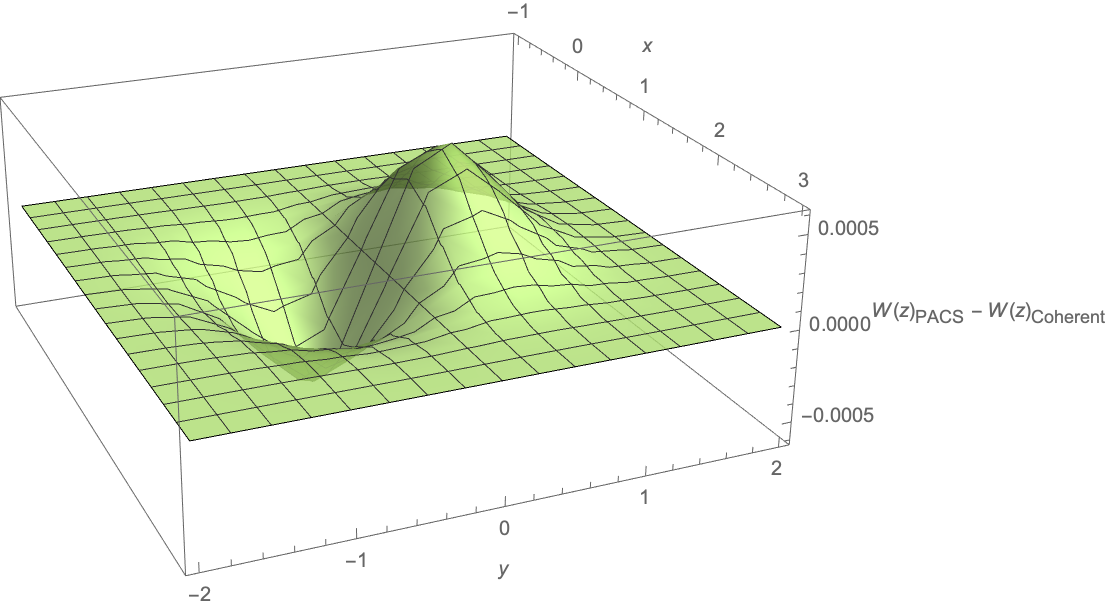}
    \caption{Difference between the Wigner function of the state \eqref{gcoherent} with $\gamma = 10^3, \abs{\alpha}=1, \mu=10^{5}\text{ Hz}, t=10^3\text{ s}, \omega = 10^{15}\text{ Hz} $ and that of a standard coherent state with $\abs{\alpha}=1$}
    \label{fig:wignernegativity}
\end{figure}

These states have been of recent interest because of their use in quantum technologies. For instance, it has been shown that they are necessary to implement a universal continuous variable quantum computer \cite{PhysRevLett.82.1784}. It is also argued that these states are required to achieve a quantum computational advantage \cite{PhysRevLett.109.230503}. Furthermore, non-gaussian states are indispensable in the implementation of various quantum-information theoretic protocols such as entanglement distillation \cite{PhysRevLett.89.137903} (see \cite{PRXQuantum.2.030204} for a detailed review on non-gaussian states). However, non-gaussian states are notoriously hard to produce and come with many practical problems \cite{PRXQuantum.2.030204}. The fact that QG effects might naturally give rise to these states is encouraging and it is therefore in our interest to probe the states we obtain in equations \eqref{gcoherent} and \eqref{ecoherent}.One way to do this is to reconstruct its Wigner function.\\ 
The Wigner function is a phase space representation of a quantum state given by a quasi-probability distribution and contains complete information about the state. It was first given by E. Wigner in 1932 \cite{PhysRev.40.749} and first measured by Smithey et al. in 1993 using optical homodyne tomography \cite{smithey1993measurement}. For quantum states that have no classical analog, the Wigner function takes on negative values, which have also been measured in experiments \cite{kurtsiefer1997measurement, *lvovsky2001quantum, *laiho2010probing}. In fact, non-gaussian states are commonly characterized by the negativity of their Wigner function\footnote{It may be noted that the negativity of Wigner function is not a necessary condition for non-classicality. For instance, in the case of the states in \eqref{gcoherent}, the Wigner function is positive everywhere due to the dominating contribution from $\ket{\tilde{\alpha}e^{i\mu t}}$. Nonetheless, the whole state is clearly not classical due to the presence of the photon-added coherent states.}\cite{kenfack2004negativity}. \\
As far as pure photon-added coherent states are concerned, they were first generated by Zavatta et al. (2004) through parametric down-conversion \cite{zavatta2004quantum}. Using balanced homodyne detection, they were able to completely characterize the states by quantum tomography and demonstrate the negativity of the Wigner function. In 2007, Zavatta et al. \cite{zavatta2007experimental} further reported the realization of single-photon-added thermal light states. In this work, they reported negative values of the Wigner function as high as of the order of $-10^{-2}$ Hz$\cdot$J$^{-1}$. Note that these experiments were purely optical and involved no light-matter interaction. Additionally, as an alternative to homodyne detection that requires one to make a series of measurements, photon-number resolving detectors could instead be used to directly measure the Wigner function \cite{lita2008counting, nehra2019state}. However, to the best of the authors' knowledge, this has never been done for photon-added coherent states.\\
In line with these experimental techniques and standards, we propose testing the production of photon-added coherent states in \eqref{gcoherent} by detecting the \textit{change} in value of the Wigner function from that of the standard coherent states that are typically produced in place of \eqref{gcoherent}. By recognizing that the change in the Wigner function is of  $\mathcal{O}(\alpha\phi\mu t)$ (see \eqref{gcoherent}), we estimate the change in the value of the Wigner function (see figure \ref{fig:wignernegativity}) to be 
$\sim 10^{-4}$ Hz$\cdot$J$^{-1}$ for $\gamma= 10^{3}$ (electroweak scale), assuming the same values for the other parameters as given in figure \ref{fig:wignernegativity}. 
Precision measurement required to measure this, say at the electroweak scale is $\Delta \text{W}(z)/\text{W}(z) \approx 10^{-4}/10^{-1}=10^{-3}$, where $\Delta \text{W}(z)$ is the change in the value of the Wigner function due to GUP corrections. Experiments in the past have achieved precision within an error of $\sim 10^{-2}$ \cite{baune2017negative, mcconnell2015entanglement, bertet2002direct}. Therefore, establishing the proposed change in the Wigner function 
should be possible 
in the near future, thereby making it feasible to test the GUP.
If these states are detected, it would indicate an intermediate length scale (of $\gamma_{0}^2l_{Pl}$) at which QG effects manifest. However, in case the states are not observed in the experiments, this method of implementing the JCM for large detuning would be useful to put improved bounds on the GUP parameter $\gamma$.

\section{Discussion and Outlook}
\label{discussion}
In this work, we estimated potentially measurable QG signatures in quantum optical systems. 
To this end, we have studied GUP effects in the JCM. Using the GUP model in \cite{PhysRevD.96.066008}, we began by computing the GUP corrected Hamiltonian. In the process, we found terms in the Hamiltonian that were quadratic and cubic in the annihilation and creation operators. We were able to neglect these terms at resonance by making the RWA (see Appendix \ref{appendix1} for details). We noted that the resultant interaction Hamiltonian \eqref{fullhamiltonian} did not have any LGUP terms since the LGUP factor $\gamma$ exclusively coupled to $\tilde{a}^2\text{ and }\tilde{a}^{\dag 2}$, which were neglected in the process of making the RWA. This happens because $\tilde{a}^2\text{ and }\tilde{a}^{\dag 2}$ terms give rise to non-energy conserving processes (such as $\ket{e,\tilde{n}}\xrightarrow[]{}\ket{g, \tilde{n+2}}$) leading to small coefficients of the resulting wave function. Moreover, this smallness exceeds the smallness from the extra GUP factor in $\gamma^2$ and thus can be ignored in comparison to the QGUP term we finally obtain in the interaction part of Hamiltonian \eqref{fullhamiltonian}.\\
After solving the obtained GUP-corrected Hamiltonian, we found the modified Rabi frequency to $\mathcal{O}(\gamma^2)$. The change in the Rabi frequency due to GUP turned out to be extremely small to be measurable.
Interestingly, on solving the JCM for the case of large detuning, we found that photon-added coherent states are created for short but experimentally feasible times. We evaluated the change in the value of the Wigner function of the resultant state \eqref{gcoherent} from that of a standard coherent state to be $\sim 10^{-4}$ Hz$\cdot$ J $^{-1}$ for a GUP perturbation at the electroweak scale. The required precision to make this measurement is calculated to be about $10^{-3}$. Noting that the recent experiments have made measurements within an error of $\sim 10^{-2}$, we 
showed that being able to measure this change will potentially test the GUP and thereby estimate QG effects in the future.
This would also be a strong piece of evidence in favour of the existence of an intermediate length scale at which QG effects manifest.
On the other hand, if these effects are smaller than we estimate, this would help narrow down the range of possible values for the GUP parameter $\gamma$, 
as experimental accuracies improve.
Either way, our results would provide useful information in the formulation and testing of QG theories. 
By providing methods for manipulation of a two-level system (or a qubit) using light, the JCM forms a link between quantum optics and quantum information theory. Jaynes Cummings-based models have been used to implement various quantum computing algorithms and quantum information-based protocols \cite{nielsen_chuang_2010}. Our work extends this utility of the JCM by showing the possibility of creating photon-added coherent states, that can further be used to implement various quantum computing and information protocols. 
Similar claims were presented in \cite{PRXQuantum.2.010325}, where the authors argue that optical interactions in the presence of QG lead to non-gaussianity. In our work we have demonstrated this explicitly by establishing an interesting link between QG and photon-added coherent states, which to the best of the authors knowledge, was not demonstrated earlier.

However, the question of whether this connection is useful is deeper and requires more sophisticated tools for answering, possibly from candidate theories of QG.

\section{Acknowledgment}

This work was supported by the Natural Sciences and Engineering Research Council of Canada and the 
Alberta Government Quantum Major Innovations Fund.

\appendix
\section{Full JCM Hamiltonian - Perturbative Treatment}
\label{appendix:perturbative}
\label{appendix1}
In the calculations done in sections \ref{correctionsquadratic} and \ref{largedetuning}, we have neglected the quadratic and cubic terms  (in $\tilde{a}$ ($\tilde{a}^\dag$)) in \eqref{modifiedinteractionhamiltonian} by making use of the RWA. We have also neglected the terms (\ref{w0+w}) and (\ref{-w0+w})
for the same reason. However, as stated towards the 
end of the section \ref{correctionsquadratic}, it is not obvious per se whether making the RWA is justified. In this appendix, we test the validity of the RWA by comparing the contributions from relevant terms in the Hamiltonian. For this, we begin by considering the full interaction Hamiltonian \eqref{modifiedinteractionhamiltonian}. We drop the term proportional to $\gamma^2\tilde{a}^{3\dag}(\tilde{a}^3)$ since these can anyway be safely neglected as pointed out in section \ref{correctionsquadratic}. Our resultant Hamiltonian then becomes
\begin{multline}
\hat{H}_{I} =  \hbar\lambda[\hat{\sigma}_{+}\{\tilde{a}^{\dag} + \tilde{a}-\phi\tilde{a}\tilde{N} + \xi\tilde{a}^2\} + \\\hat{\sigma}_{-}\{\tilde{a}+\tilde{a}^{\dag} - \phi\tilde{a}^{\dag}(\tilde{N}+1) - \xi\tilde{a}^{\dag 2}\}].
\end{multline}
where we have defined $\xi = i\delta\gamma\sqrt{2\hbar\omega}$. In the above Hamiltonian, $\phi$ denotes QGUP corrections whereas $\xi$ denotes LGUP corrections. 
The terms that we need to account for are those that are proportional to $\hat{\sigma}_{+}\tilde{a}^{\dag}$ ($\hat{\sigma}_{-}\tilde{a}$), and terms that are quadratic in the annihilation and creation operators. Note that the latter are precisely the terms linear in the GUP parameter $\gamma$. Therefore, showing that the contribution of the terms that are quadratic in $\tilde{a}(\tilde{a}^{\dag})$ is small automatically justifies our restriction to the QGUP model. Since an analytic solution is not easily found, we assume the interaction strength $\lambda$ to be small and perform a perturbative study of the dynamics of the system. 
We begin by assuming the initial state to be $\ket{\psi(0)} = \ket{e}\ket{n}$. Up to first order in $\lambda$, a general ansatz for $\ket{\psi(t)}$ can be written as \cite{shankar2012principles}, 

\begin{multline}
    \ket{\psi(t)} = C^{(1)}_{g,n-1}(t)\ket{g}\ket{n-1}e^{-E_{g,n-1}\frac{it}{\hbar}}+ \\C^{(1)}_{e,n}(t)\ket{e}\ket{n}e^{-E_{e,n}\frac{it}{\hbar}}+ C^{(1)}_{g,n+1}(t)\ket{g}\ket{n+1}e^{-E_{g,n+1}\frac{it}{\hbar}}\\ + C^{(1)}_{g,n+2}(t)\ket{g}\ket{n+2}e^{-E_{g,n+2}\frac{it}{\hbar}}
\end{multline}
where $\{E_{e,n}, E_{g,n}\}$ and $\{C_{e,n}, C_{g,n}\}$ denote the energies and amplitudes of the states $\{\ket{e,n}, \ket{g,n}\}$ respectively. The amplitudes can be calculated using standard first-order perturbation theory.

\begin{figure}
    \centering
    \includegraphics[width =0.5\textwidth]{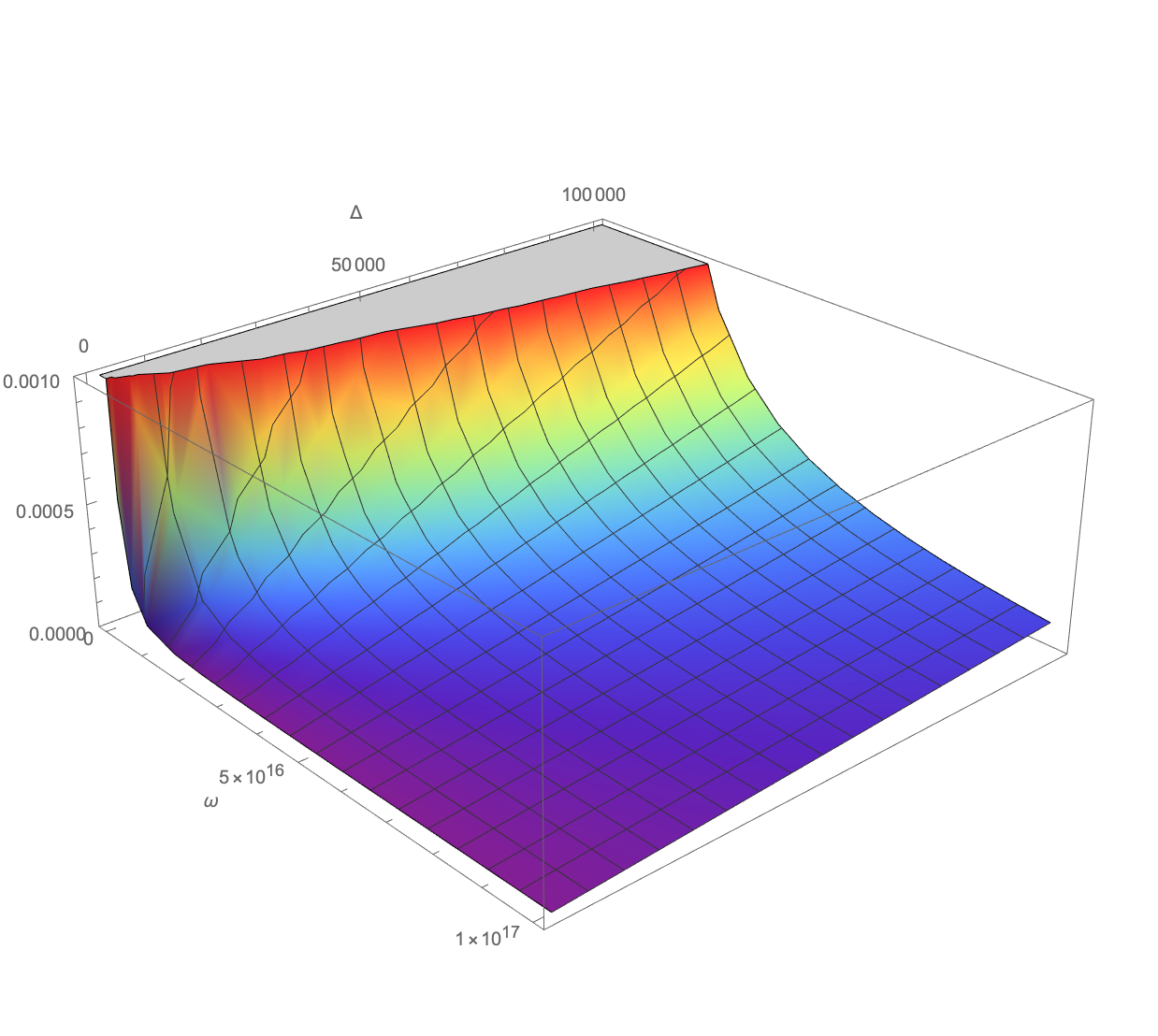}
    \caption{$\zeta_{LQ} \text{ for } n=50, \gamma = 0.5, \delta=1, \epsilon=1 \text{ as a function of } \omega \text{ and } \Delta = \omega_{0}-\omega$}
    \label{zetalq}
\end{figure}

Since there is no transition from $\ket{e}\ket{n}$ $\xrightarrow[]{}$ $\ket{e}\ket{n}$ via an intermediate state (to first order in $\lambda$), we have $C^{(1)}_{e,n}(t)=0$. Evaluating the other amplitudes to $\mathcal{O}(\lambda)$ gives:
\begin{gather}
\label{cgn-1}
C^{(1)}_{g,n-1}(t) = \lambda\sqrt{n}\frac{e^{-i(\omega+\omega_0)t} - 1}{\omega+\omega_0} \\
\begin{split}
\label{cgn+1}
 C^{(1)}_{g,n+1}(t) =-\lambda\sqrt{(n+1)}\bigg(1 - (n+1)\phi\bigg)
\bigg(\frac{e^{i(\omega - \omega_0)t}-1}{\omega-\omega_0}\bigg)
\end{split}\\
\label{cgn+2}
 C^{(1)}_{g,n+2}(t) =\lambda\xi\sqrt{(n+1)(n+2)}
\bigg(\frac{e^{i(2\omega - \omega_0)t}-1}{2\omega-\omega_0}\bigg)~.
\end{gather}
As can be seen, the RHS of the above equations 
have a number of parameters, namely 
$\omega, \omega_{0}, \gamma, \delta, \epsilon, \text{ and}, \lambda$. It is instructive to compare the contribution of various terms by assuming a range of operational values for the parameters.
To do this, it is first useful to take the absolute values of time averages of the expressions \eqref{cgn-1}-\eqref{cgn+2}, where we denote the averages by an overbar. This gives
\begin{gather}
\label{cgn-1abs}
\abs{\overline{C^{(1)}_{g,n-1}(t)}} = \lambda\frac{\sqrt{n}}{\omega+\omega_0} \\
\begin{split}
\label{cgn+1abs}
 \abs{\overline{C^{(1)}_{g,n+1}(t)}} =&\lambda\bigg(\sqrt{(n+1)} - \\&(n+1)^{3/2}\hbar\omega\gamma^2(3\delta^2-2\epsilon)\bigg)
\frac{1}{\omega-\omega_0}
\end{split}\\
\label{cgn+2abs}
 \abs{\overline{C^{(1)}_{g,n+2}(t)}} =\lambda\delta\gamma\sqrt{2\hbar\omega}\sqrt{(n+1)(n+2)}
\bigg(\frac{1}{2\omega-\omega_0}\bigg)
\end{gather}
where we have expanded $\xi$ and $\phi$. In equation \eqref{cgn+1abs}, we note that only the second term is of interest to us since that is the part with the relevant GUP factor. We denote this term as $\text{T}_{2}(C^{(1)}_{g,n+1}(t))$. The first term signifies the emission of a photon in the standard JCM and is not relevant for the purposes of this comparison. We weigh the contribution of various terms by taking the ratios of the above expressions. To this end, we define
\begin{equation}
    \zeta_{LQ} = \frac{\abs{\overline{C^{(1)}_{g,n+2}(t)}}}{\text{T}_{2}(C^{(1)}_{g,n+1}(t))}
\end{equation}
and 
\begin{equation}
 \zeta_{RQ} = \frac{\abs{\overline{C^{(1)}_{g,n-1}(t)}}}{\text{T}_{2}(C^{(1)}_{g,n+1}(t))}.
\end{equation}
$\zeta_{LQ}$ signifies the ratio of strengths of the of the LGUP to the QGUP terms, whereas 
$\zeta_{RQ}$ signifies the ratio of the strengths of the terms $\hat{\sigma}_{+}\tilde{a}^{\dag}$ ($\hat{\sigma}_{-}\tilde{a}$) (that are typically neglected as part of the RWA in standard JCM) to the QGUP terms.
We begin by first considering $\zeta_{LQ}$. Evaluating, we get
\begin{equation}
\zeta_{LQ} = \frac{\sqrt{2(n+2)}}{n+1}\frac{\delta}{3\delta^2 - 2\epsilon}\frac{1}{\gamma\sqrt{\hbar\omega}}\frac{\omega-\omega_0}{2\omega-\omega_0}
\end{equation}
We assume certain experimentally viable values for the above parameters to estimate this ratio. Taking $n\in(1, 100), \;\gamma \in (10^{-5}, 10^3),\; \omega_{0}-\omega = \Delta \in (10^{3}, 10^5), \omega \in (10^9, 10^{17}),\text{ and }(\delta,\epsilon)\in(0, 1)$, one can easily see that there exist realizable values of the parameters in which one could neglect the LGUP contribution, i.e., terms that are quadratic in $\tilde{a}(\tilde{a}^{\dag})$. For instance, if light frequencies exceed $10^{16}\text{ Hz }$, one can always neglect the contribution of the LGUP terms in comparison to the QGUP terms, assuming $n=50$ and $\delta=\epsilon=1$ (see figure \ref{zetalq}). %\textcolor{red}
However, this true only if $\gamma\geq10^{-1}$, which translates to a length scale greater than $10^{-27}$m. If GUP takes effect at lower length scales, our simulations show that the LGUP contribution to the dynamics cannot be neglected assuming working values of the other parameters.
\\
Next, consider $\zeta_{RQ}$. Evaluating, we get 
\begin{equation}
    \zeta_{RQ} = \frac{\sqrt{n}}{(n+1)^{3/2}}\frac{\omega-\omega_0}{\omega+\omega_0}\frac{1}{3\delta^2 - 2\epsilon}\frac{1}{\gamma^2}\frac{1}{\hbar\omega}
\end{equation}
Keeping the same range of the parameters, we find that it is possible to find regimes wherein contributions of the $\hat{\sigma}_{+}\tilde{a}^{\dag}$ ($\hat{\sigma}_{-}\tilde{a}$) terms can be neglected in comparison to the QGUP contributions. 
\begin{figure}
    \centering
    \includegraphics[width = 0.5\textwidth]{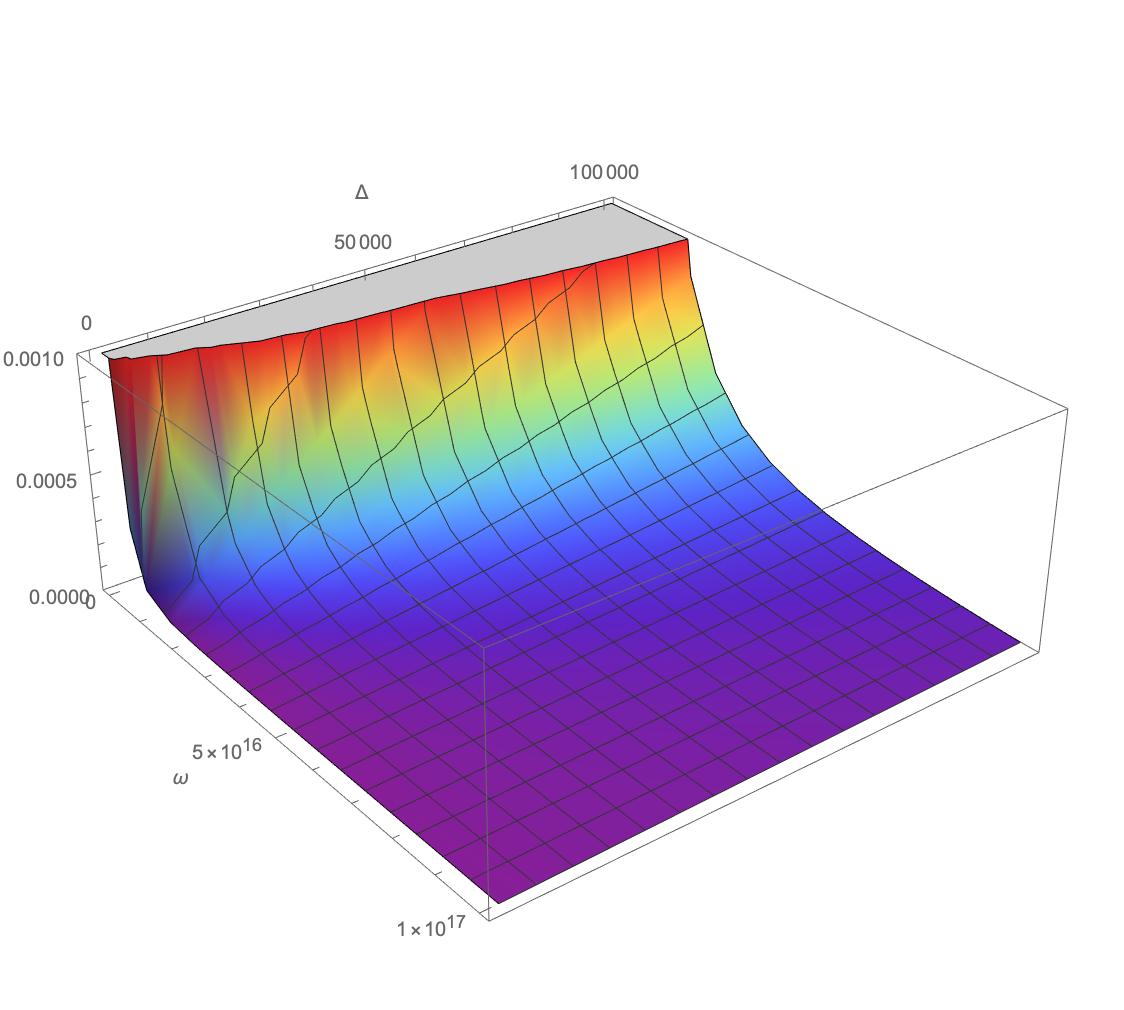}
    \caption{$\zeta_{RQ} \text{ for } n=50, \gamma = 5\times10^{3}, \delta=1, \epsilon=1 \text{ as a function of } \omega \text{ and } \Delta = \omega_{0}-\omega$}
    \label{zetarq}
\end{figure}
In figure \ref{zetarq} for instance, we see that this can be done for light frequencies greater than $10^{16}$ Hz. 
However, this is only true if $\gamma=10^{3}$, which translates to around the electroweak length scale of about $10^{-18}$m. If $\gamma$ is below this threshold, or if GUP takes effect below this length scale, we find that it gets increasingly difficult to find experimentally realizable regimes where this approximation can be made.\\%} 
Thus, we see that neglecting the contributions of $\hat{\sigma}_{+}\tilde{a}^{\dag}$ ($\hat{\sigma}_{-}\tilde{a}$) and LGUP terms to the dynamics is a reasonable assumption after all, since this can be done for a range of parameter values, many of which are experimentally viable. 

\section{Derivation of the Effective Action}
\label{appendix:effectivaction}
We briefly present the derivation of the effective Hamiltonian for an interaction with large detuning $\Delta = \omega_{0}-\omega$ (considering only the QGUP corrections). The derivation closely follows the derivation in Appendix C of \cite{gerry_knight_2004}. The difference is that we have a few extra terms that come into the Hamitlonian because of the GUP. We begin by considering the full Hamiltonian
\begin{equation}
\hat{H} = \hat{H}_{0} + \hat{H}_{I}
\end{equation}
where $\hat{H}_{0} = \frac{1}{2}\hbar\omega_{0}\hat{\sigma}_{3} + \hbar\omega\big[\tilde{N}-\{4(\tilde{N}^{2}+\tilde{N})\chi + \beta\}  \big]$ is the interaction-free Hamiltonian (from \ref{fullhamiltonian}) and $\hat{H}_{I} = \hbar\lambda\{\hat{\sigma}_{+}(\tilde{a}-\tilde{a}\tilde{N}\phi) + \hat{\sigma}_{-}(\tilde{a}^{\dag} - \tilde{a}^{\dag}(\tilde{N}+1)\phi)\}$ is the interacting Hamiltonian. We can write $\hat{H}_{I}$ in the form 
\begin{equation}
    \hat{H}_{I} = \hbar\lambda({\hat{A}+\hat{A}^{\dag}})
\end{equation}
where $\hat{A} = \hat{\sigma}_{+}\tilde{a}(1-\tilde{N}\phi)$. Transforming to the interaction picture (IP), the Hamiltonian is given as
\begin{equation}
    \hat{H}_{IP} = \hat{U}^{-1}_{0}\hat{H}\hat{U}_{0} - i\hbar\hat{U}^{-1}_{0}\frac{d\hat{U}_{0}}{dt}
\end{equation}
where $U_{0} = \exp(-iH_{0}t/\hbar)$. Computing this for the above Hamiltonian, we get
\begin{equation}
\hat{H}_{IP} = \hbar\lambda(\hat{A}e^{it(\Delta+8n\chi\omega)} + \hat{A}^{\dag}e^{-it(\Delta+8\chi\omega(n+1))})
\end{equation}
where $\Delta = \omega_0 - \omega$ is assumed to be large. The state vector in the interaction picture obeys the Schrödinger equation with the Hamiltonian $H_{IP}$. The solution for $\ket{\psi_{IP}(t)}$ can then be written as
\begin{equation}
    \ket{\psi_{IP}(t)} = \hat{\mathcal{T}}\bigg[\exp\bigg(\frac{-i}{\hbar}\int_{0}^{t}dt'\hat{H}_{IP}(t')\bigg)\bigg]\ket{\psi_{IP}(0)}
\end{equation}
where $\ket{\psi_{IP}(t)}$ is the state-vector in the interaction picture at time $t$ and $\hat{\mathcal{T}}$ is the time-ordering operator. Making the perturbative expansion, we get
\begin{multline}
    \hat{\mathcal{T}}\bigg[\exp\bigg(\frac{-i}{\hbar}\int_{0}^{t}dt'\hat{H}_{IP}(t')\bigg)\bigg] = \hat{I} - \frac{i}{\hbar}\int_{0}^{t}dt'\hat{H}_{IP}(t')\\ - \frac{1}{2\hbar^2}\hat{\mathcal{T}}\bigg[\int_{0}^{t}dt'\int_{0}^{t}dt''\hat{H}_{IP}(t')\hat{H}_{IP}(t'') \bigg] + ...
\end{multline}
Evaluating the above and ignoring terms that are of ($\mathcal{O}(\lambda^2/\Delta^2)$), we get
\begin{multline}
\label{effectivehamiltonianexpansion}
 \hat{\mathcal{T}}\bigg[\exp\bigg(\frac{-i}{\hbar}\int_{0}^{t}dt'\hat{H}_{IP}(t')\bigg)\bigg] = \hat{I} - \lambda\bigg[\hat{A}\frac{(e^{it(\Delta+8n\omega\chi)}- 1)}{\Delta+8n\omega\chi}\\- \hat{A}^{\dag}\frac{(e^{-it(\Delta+8\omega\chi(n+1))}- 1)}{\Delta+8\omega\chi(n+1)}\bigg] - \frac{it\lambda^2}{\Delta}[\hat{A}, \hat{A}^{\dag}]
\end{multline}
Note that we have only kept the terms up to $\mathcal{O}(\gamma^2) (\mathcal{O}(\chi))$ in the final term above. Now, if the mean quanta in the light-field $\langle\hat{A}^{\dag}\hat{A}\rangle^{1/2}$ is not too large and if
\begin{equation}
    \abs{\frac{\lambda}{\Delta}\langle\hat{A}^{\dag}\hat{A}\rangle^{1/2}}\ll 1
\end{equation}
due to the large detuning ($\Delta$), we can drop the second term in equation \ref{effectivehamiltonianexpansion}. Finally, we get
\begin{equation}
     \hat{\mathcal{T}}\bigg[\exp\bigg(\frac{-i}{\hbar}\int_{0}^{t}dt'\hat{H}_{IP}(t')\bigg)\bigg] \approx \hat{I} - \frac{it}{\hbar}\hat{H}_{\text{eff}},
\end{equation}
where 
\begin{equation}
    \hat{H}_{\text{eff}} = \frac{\hbar \lambda^2}{\Delta}[\hat{A}, \hat{A}^{\dag}]
\end{equation}
is the effective Hamiltonian. Note that the above expression is well in accordance with the expression derived in \cite{gerry_knight_2004} if we let the GUP parameter $\gamma$ (written in terms of $\chi$ in the above expressions) to be 0. 
\bibliographystyle{unsrt}
\bibliography{bibliography.bib}

\begin{thebibliography}{10}

\bibitem{PhysRevA.43.492}
G.~S. Agarwal and K.~Tara.
\newblock Nonclassical properties of states generated by the excitations on a
  coherent state.
\newblock {\em Phys. Rev. A}, 43:492--497, Jan 1991.

\bibitem{PhysRevD.84.044013}
Ahmed~Farag Ali, Saurya Das, and Elias~C. Vagenas.
\newblock Proposal for testing quantum gravity in the lab.
\newblock {\em Phys. Rev. D}, 84:044013, Aug 2011.

\bibitem{PhysRevLett.101.221301}
Saurya Das and Elias~C. Vagenas.
\newblock Universality of quantum gravity corrections.
\newblock {\em Phys. Rev. Lett.}, 101:221301, Nov 2008.

\bibitem{ALI2009497}
Ahmed~Farag Ali, Saurya Das, and Elias~C. Vagenas.
\newblock Discreteness of space from the generalized uncertainty principle.
\newblock {\em Physics Letters B}, 678(5):497--499, 2009.

\bibitem{osti_22701519}
Pasquale Bosso and Saurya Das.
\newblock Generalized uncertainty principle and angular momentum.
\newblock {\em Annals of Physics}, 383, 8 2017.

\bibitem{shababi2020non}
Homa Shababi and Kamel Ourabah.
\newblock Non-gaussian statistics from the generalized uncertainty principle.
\newblock {\em The European Physical Journal Plus}, 135(9):1--13, 2020.

\bibitem{aghababaei2022quantum}
Sarah Aghababaei, Hooman Moradpour, and Hamid Shabani.
\newblock Quantum gravity and the square of bell operators.
\newblock {\em Quantum Information Processing}, 21(2):1--10, 2022.

\bibitem{lemos2022quantum}
AS~Lemos and FA~Brito.
\newblock Quantum gravity constraints on fine structure constant from gup in
  braneworlds.
\newblock {\em arXiv preprint arXiv:2203.11403}, 2022.

\bibitem{bosso2022bell}
Pasquale Bosso, Luciano Petruzziello, Fabian Wagner, and Fabrizio Illuminati.
\newblock Bell nonlocality in quantum-gravity induced minimal-length quantum
  mechanics.
\newblock {\em arXiv preprint arXiv:2207.10418}, 2022.

\bibitem{Pikovski2012Probing}
Igor Pikovski, Michael~R. Vanner, Markus Aspelmeyer, M.~S. Kim, and Caslav
  Brukner.
\newblock {Probing Planck-scale physics with quantum optics}.
\newblock {\em Nature Physics}, 8(5):393--397, March 2012.

\bibitem{PhysRevA.96.023849}
Pasquale Bosso, Saurya Das, Igor Pikovski, and Michael~R. Vanner.
\newblock Amplified transduction of planck-scale effects using quantum optics.
\newblock {\em Phys. Rev. A}, 96:023849, Aug 2017.

\bibitem{Marin:2013pga}
Francesco Marin et~al.
\newblock {Gravitational bar detectors set limits to Planck-scale physics on
  macroscopic variables}.
\newblock {\em Nature Phys.}, 9:71--73, 2013.

\bibitem{PhysRevA.90.033834}
Andreas Albrecht, Alex Retzker, and Martin~B. Plenio.
\newblock Testing quantum gravity by nanodiamond interferometry with
  nitrogen-vacancy centers.
\newblock {\em Phys. Rev. A}, 90:033834, Sep 2014.

\bibitem{2015}
Mateusz Bawaj, Ciro Biancofiore, Michele Bonaldi, Federica Bonfigli, Antonio
  Borrielli, Giovanni Di~Giuseppe, Lorenzo Marconi, Francesco Marino, Riccardo
  Natali, Antonio Pontin, and et~al.
\newblock Probing deformed commutators with macroscopic harmonic oscillators.
\newblock {\em Nature Communications}, 6(1), Jun 2015.

\bibitem{1443594}
E.T. Jaynes and F.W. Cummings.
\newblock Comparison of quantum and semiclassical radiation theories with
  application to the beam maser.
\newblock {\em Proceedings of the IEEE}, 51(1):89--109, 1963.

\bibitem{gerry_knight_2004}
Christopher Gerry and Peter Knight.
\newblock {\em Introductory Quantum Optics}.
\newblock Cambridge University Press, 2004.

\bibitem{PhysRevLett.103.086403}
S.~Schmidt and G.~Blatter.
\newblock Strong coupling theory for the jaynes-cummings-hubbard model.
\newblock {\em Phys. Rev. Lett.}, 103:086403, Aug 2009.

\bibitem{nielsen_chuang_2010}
Michael~A. Nielsen and Isaac~L. Chuang.
\newblock {\em Quantum Computation and Quantum Information: 10th Anniversary
  Edition}.
\newblock Cambridge University Press, 2010.

\bibitem{azuma2011quantum}
Hiroo Azuma.
\newblock Quantum computation with the jaynes-cummings model.
\newblock {\em Progress of Theoretical Physics}, 126(3):369--385, 2011.

\bibitem{PhysRevA.69.062320}
Alexandre Blais, Ren-Shou Huang, Andreas Wallraff, S.~M. Girvin, and R.~J.
  Schoelkopf.
\newblock Cavity quantum electrodynamics for superconducting electrical
  circuits: An architecture for quantum computation.
\newblock {\em Phys. Rev. A}, 69:062320, Jun 2004.

\bibitem{2020}
Valentin Kasper, Gediminas Juzeli{\={u}}nas, Maciej Lewenstein, Fred
  Jendrzejewski, and Erez Zohar.
\newblock From the jaynes{\textendash}cummings model to non-abelian gauge
  theories: a guided tour for the quantum engineer.
\newblock 22(10):103027, oct 2020.

\bibitem{greentree2013fifty}
Andrew~D Greentree, Jens Koch, and Jonas Larson.
\newblock Fifty years of jaynes--cummings physics.
\newblock {\em Journal of Physics B: Atomic, Molecular and Optical Physics},
  46(22):220201, 2013.

\bibitem{PhysRevD.96.066008}
Pasquale Bosso, Saurya Das, and Robert~B. Mann.
\newblock Planck scale corrections to the harmonic oscillator, coherent, and
  squeezed states.
\newblock {\em Phys. Rev. D}, 96:066008, Sep 2017.

\bibitem{bosso2018potential}
Pasquale Bosso, Saurya Das, and Robert~B Mann.
\newblock Potential tests of the generalized uncertainty principle in the
  advanced ligo experiment.
\newblock {\em Physics Letters B}, 785:498--505, 2018.

\bibitem{PhysRevD.52.1108}
Achim Kempf, Gianpiero Mangano, and Robert~B. Mann.
\newblock Hilbert space representation of the minimal length uncertainty
  relation.
\newblock {\em Phys. Rev. D}, 52:1108--1118, Jul 1995.

\bibitem{doi:10.1080/09500349314551321}
Bruce~W. Shore and Peter~L. Knight.
\newblock The jaynes-cummings model.
\newblock {\em Journal of Modern Optics}, 40(7):1195--1238, 1993.

\bibitem{brune1994lamb}
M~Brune, P~Nussenzveig, F~Schmidt-Kaler, F~Bernardot, Abdelhamid Maali,
  JM~Raimond, and S~Haroche.
\newblock From lamb shift to light shifts: Vacuum and subphoton cavity fields
  measured by atomic phase sensitive detection.
\newblock {\em Physical review letters}, 72(21):3339, 1994.

\bibitem{guo1996generation}
Guang-Can Guo and Shi-Biao Zheng.
\newblock Generation of schr{\"o}dinger cat states via the jaynes-cummings
  model with large detuning.
\newblock {\em Physics Letters A}, 223(5):332--336, 1996.

\bibitem{PhysRevLett.82.1784}
Seth Lloyd and Samuel~L. Braunstein.
\newblock Quantum computation over continuous variables.
\newblock {\em Phys. Rev. Lett.}, 82:1784--1787, Feb 1999.

\bibitem{PhysRevLett.109.230503}
A.~Mari and J.~Eisert.
\newblock Positive wigner functions render classical simulation of quantum
  computation efficient.
\newblock {\em Phys. Rev. Lett.}, 109:230503, Dec 2012.

\bibitem{PhysRevLett.89.137903}
J.~Eisert, S.~Scheel, and M.~B. Plenio.
\newblock Distilling gaussian states with gaussian operations is impossible.
\newblock {\em Phys. Rev. Lett.}, 89:137903, Sep 2002.

\bibitem{PRXQuantum.2.030204}
Mattia Walschaers.
\newblock Non-gaussian quantum states and where to find them.
\newblock {\em PRX Quantum}, 2:030204, Sep 2021.

\bibitem{PhysRev.40.749}
E.~Wigner.
\newblock On the quantum correction for thermodynamic equilibrium.
\newblock {\em Phys. Rev.}, 40:749--759, Jun 1932.

\bibitem{smithey1993measurement}
DT~Smithey, M~Beck, Michael~G Raymer, and A~Faridani.
\newblock Measurement of the wigner distribution and the density matrix of a
  light mode using optical homodyne tomography: Application to squeezed states
  and the vacuum.
\newblock {\em Physical review letters}, 70(9):1244, 1993.

\bibitem{kurtsiefer1997measurement}
Ch~Kurtsiefer, T~Pfau, and J~Mlynek.
\newblock Measurement of the wigner function of an ensemble of helium atoms.
\newblock {\em Nature}, 386(6621):150--153, 1997.

\bibitem{lvovsky2001quantum}
Alexander~I Lvovsky, Hauke Hansen, T~Aichele, O~Benson, J~Mlynek, and
  S~Schiller.
\newblock Quantum state reconstruction of the single-photon fock state.
\newblock {\em Physical Review Letters}, 87(5):050402, 2001.

\bibitem{laiho2010probing}
Kaisa Laiho, Kati{\'u}scia~N Cassemiro, David Gross, and Christine Silberhorn.
\newblock Probing the negative wigner function of a pulsed single photon point
  by point.
\newblock {\em Physical review letters}, 105(25):253603, 2010.

\bibitem{kenfack2004negativity}
Anatole Kenfack and Karol {\.Z}yczkowski.
\newblock Negativity of the wigner function as an indicator of
  non-classicality.
\newblock {\em Journal of Optics B: Quantum and Semiclassical Optics},
  6(10):396, 2004.

\bibitem{zavatta2004quantum}
Alessandro Zavatta, Silvia Viciani, and Marco Bellini.
\newblock Quantum-to-classical transition with single-photon-added coherent
  states of light.
\newblock {\em science}, 306(5696):660--662, 2004.

\bibitem{zavatta2007experimental}
Alessandro Zavatta, Valentina Parigi, and Marco Bellini.
\newblock Experimental nonclassicality of single-photon-added thermal light
  states.
\newblock {\em Physical Review A}, 75(5):052106, 2007.

\bibitem{lita2008counting}
Adriana~E Lita, Aaron~J Miller, and Sae~Woo Nam.
\newblock Counting near-infrared single-photons with 95\% efficiency.
\newblock {\em Optics express}, 16(5):3032--3040, 2008.

\bibitem{nehra2019state}
Rajveer Nehra, Aye Win, Miller Eaton, Reihaneh Shahrokhshahi, Niranjan Sridhar,
  Thomas Gerrits, Adriana Lita, Sae~Woo Nam, and Olivier Pfister.
\newblock State-independent quantum state tomography by photon-number-resolving
  measurements.
\newblock {\em Optica}, 6(10):1356--1360, 2019.

\bibitem{baune2017negative}
Christoph Baune, Jarom{\'\i}r Fiur{\'a}{\v{s}}ek, and Roman Schnabel.
\newblock Negative wigner function at telecommunication wavelength from
  homodyne detection.
\newblock {\em Physical Review A}, 95(6):061802, 2017.

\bibitem{mcconnell2015entanglement}
Robert McConnell, Hao Zhang, Jiazhong Hu, Senka {\'C}uk, and Vladan
  Vuleti{\'c}.
\newblock Entanglement with negative wigner function of almost 3,000 atoms
  heralded by one photon.
\newblock {\em Nature}, 519(7544):439--442, 2015.

\bibitem{bertet2002direct}
Patrice Bertet, Alexia Auffeves, Paolo Maioli, Stefano Osnaghi, Tristan
  Meunier, Michel Brune, Jean-Michel Raimond, and Serge Haroche.
\newblock Direct measurement of the wigner function of a one-photon fock state
  in a cavity.
\newblock {\em Physical Review Letters}, 89(20):200402, 2002.

\bibitem{PRXQuantum.2.010325}
Richard Howl, Vlatko Vedral, Devang Naik, Marios Christodoulou, Carlo Rovelli,
  and Aditya Iyer.
\newblock Non-gaussianity as a signature of a quantum theory of gravity.
\newblock {\em PRX Quantum}, 2:010325, Feb 2021.

\bibitem{shankar2012principles}
Ramamurti Shankar.
\newblock {\em Principles of quantum mechanics}.
\newblock Springer Science \& Business Media, 2012.

\end{thebibliography}

\end{document}